# Systematic of Particle Thermal Freeze-out in a Hadronic Fireball at RHIC

Saeed Uddin[*], Riyaz Ahmed Bhat, Inam-ul Bashir, Waseem Bashir

*Department of Physics, Jamia Millia Islamia (Central University)*

*New Delhi-110025*

and

Jan Shabir Ahmad

*Department of Physics, Govt. Degree College, Baramulla*

*Srinagar, J & K.*

## Abstract

We attempt to describe the rapidity and transverse momentum spectra of strange as well as non-strange hadrons e.g. $\Xi, \bar{\Xi}, \Lambda, \bar{\Lambda}, p, \bar{p}, (\Omega^- + \Omega^+), K^+, K^-$ and their ratios in the ultra-relativistic collisions of gold nuclei at $\sqrt{s_{NN}} = 200$ GeV. This is done by using a statistical thermal freeze-out model which incorporates the rapidity (collision) axis as well as transverse direction boosts developed within an expanding hot and dense hadronic fluid (fireball) till the final freeze-out. We determine the thermo-chemical freeze-out conditions particularly in terms of the temperature, baryon chemical potential and collective flow effect parameters for different particle species. The parameters indicate occurrence of freeze-out of the singly and doubly strange hyperon species at somewhat earlier times during the evolution of the fireball. The experimental data of the transverse momentum and rapidity distribution are well reproduced. The contribution of heavier hadronic resonance decay is taken into account.

[*] *E mail : saeed_jmi@yahoo.co.in*

## 1. Introduction

It is commonly believed that in the ultra-relativistic nuclear collisions a hot fireball is formed in a very short time (~ 1 fm/c) after the initial hard parton collisions of nuclei [1-2]. If the initial temperature of a fireball $T_i$ is high, then it may be in a deconfined state of quarks and gluons, called Quark Gluon Plasma (QGP). If this state of matter last for a sufficient period of time ($\tau \sim$ a few fm/c) then a locally thermally and chemically equilibrated QGP phase is likely to exist. As the initial number density of the produced partons is very large in such collisions therefore the mean free path of these partons is extremely small leading to multiple collisions among them. Within the framework of the statistical model it is assumed that this hot fireball undergoes expansion accompanied by particle production processes which consequently leads to a decrease in its temperature. At certain critical temperature, also called phase transition temperature, $T_c$, quarks and gluons start forming hadrons (i.e. hadronization via recombination) and the system first goes in a mixed state. When the temperature $T < T_c$, the matter is entirely composed of various hadronic species which continue to interact inelastically causing secondary particle (hadron) production leading to a further change in the particle numbers. At certain stage the particle number changing (inelastic) processes stop and a hadro-chemical or simply a chemical freeze-out occurs. The constituent particles in the dense hadronic matter still continue to interact elastically causing a fluid like expansion of the matter leading to a further drop in the temperature of the system till the final thermal/hydrodynamical freeze-out [3-5]. After this stage the system will ultimately disintegrate into the final state of non-interacting individual hadrons. Hence it is clear that if the system lasts long enough ($\tau \sim 10$ few fm/c) it is likely to reach a final state having a high degree of chemical as well as thermal equilibrium at freeze-out. This can be described by nearly free (non interacting) gas of various hadronic resonances. This final freeze-out essentially occurs when the rate of microscopic hadronic interactions becomes comparable to that of macroscopic expansion of the system [6]. The hydrodynamic expansion is important as it leads to the reshaping of the particle spectra.

As the yields of baryons and anti-baryons are an important indicator of the multi-particle production phenomenon in the ultra-relativistic nuclear collisions hence their distribution allows us to learn how the baryon number of the system, which is initially carried by the nucleons only, before the nuclear collision, is distributed at the final freeze-out state [4]. It is believed that these produced hadrons carry information about the collision dynamics and the subsequent space-time evolution of the system. Hence a precise measurement of the transverse momentum ($p_T$) and rapidity distributions of identified hadrons is essential for the understanding of the thermal and hydrodynamical properties of the created matter up to the final freeze-out. Some hydrodynamical models [2,3] that include radial flow have successfully described the measured $p_t$ distributions in Au+Au collisions at $\sqrt{s_{NN}}$= 130 GeV [4,7]. The $p_T$ spectra of identified charged hadrons below 2 GeV/c in central collisions have been well reproduced in some models by two simple parameters: transverse flow velocity $\beta_T$ and thermal freeze-out temperature, T under the assumption of thermalization [7]. Some statistical models have also successfully described the particle abundances but at low $p_T$ [8-10]. In the following, we attempt to briefly describe our model [4] and obtain the expressions for the transverse momentum and the rapidity distribution of particles at the final freeze-out.

## 2. The Model

First analysis of the data obtained by the BRAHMS collaboration at $\sqrt{s_{NN}} = 200$ GeV was done by Stiles and Murray [11]. The BRAHMS collaboration data at $\sqrt{s_{NN}} = 200$ GeV has a clear dependence of the baryon chemical potential on rapidity which is revealed through the changing $\overline{p}/p$ ratio with rapidity. The general procedure of our model is as follows: that we incorporate this effect by considering a thermal model which assumes that the rapidity axis is populated with hot regions moving along the beam axis with monotonically increasing rapidity $y_0$. This essentially emerges from the situation where the colliding nuclei exhibit (at least partial) transparency. Hence the regions of the matter away from the mid region ($z \approx 0$) also consist of the constituent partons of the colliding nucleons, which suffer less rapidity loss due to partial nuclear transparency. Thus these regions have an excess of quarks over the anti-quarks and hence maintain larger baryon chemical potentials on *either side* of the mid-rapidity region in a symmetric manner. Unlike the earlier assumption of taking temperature also to be rapidity dependent [12-16], here the temperature is assumed to be the nearly same for all the hadronic fluid elements [4]. In addition to the longitudinal expansion we also incorporate transverse expansion. At the final thermal/hydrodynamical freeze-out, which follows the chemical freeze-out, the emitted particles leave the different regions of the fireball following a local (thermal) distribution. The resulting rapidity and transverse momentum distributions of any given particle specie is then obtained by a superposition of the contributions of all these regions.

The momentum distributions of hadrons, emitted from within an expanding fireball in the state of local thermal equilibrium, are characterized by the Lorentz-invariant Cooper-Frye formula [6]

$$E\frac{d^3n}{d^3P} = \frac{g}{(2\pi)^3} \int f\left(\frac{p^\mu u^\mu}{T}, \lambda\right) p^\mu d\Sigma_\mu, \qquad (1)$$

where $\Sigma_f$ represents the freeze-out hyper-surface.

In order to obtain the particle spectra in the *overall* rest frame of the hadronic fireball (i.e. the rest frame of the colliding nuclei) in our model, we first define the invariant cross-section for given hadronic specie in the *local rest frame of a hadronic fluid element*. As the invariant cross section will have the same value in all Lorentz frames [17], we can thus write,

$$E \frac{d^3n}{d^3p} = E' \frac{d^3n}{d^3p'} \qquad (2)$$

The primed quantities on the RHS refer to the invariant spectra of given hadronic species in the rest frame of a local hadronic fluid element, while the unprimed quantities on the LHS refer to the invariant spectra of the same hadronic species but in the overall rest frame of the hadronic fireball formed in the ultra-relativistic central nuclear collisions. The occupation number distribution of the hadrons in the momentum space is assumed to follow the quantum distribution function

$$E' \frac{d^3n}{d^3p'} \sim \frac{E'}{e^{(\frac{E'-\mu}{T})} \pm 1} \qquad (3)$$

Where (+) sign and (-) sign are for fermions and bosons respectively and μ is the chemical potential of the given hadronic specie.

In recent works it has been clearly shown that there is a strong evidence of increasing baryon chemical potential, $\mu_B$ along the collision axis in the RHIC experiments [4, 18]. We can parameterize the chemical potential as a function of expanding fireball rapidity, $y_0$, as [4,5,18]

$$\mu_B = a + b y_0^2 \qquad (4)$$

Where the quadratic type dependence of the baryon chemical potential, $\mu_B$, on $y_0$ has been considered in above equation which also makes $\mu_B$ invariant under the transformation $y_0 \rightarrow -y_0$, as the system properties are to remain invariant under the above transformation in the rest frame of the fireball formed. The energy and momentum of the particles in the rest frame (primed) of the *local hadronic fluid element* in terms of the unprimed quantities are given by the Lorentz transformation as

$$E' = \gamma(E - \vec{p}.\vec{\beta}) \quad \text{with} \quad \vec{p}.\vec{\beta} = p_T \beta_T + p_z \beta_z \tag{5}$$

$$p' = (E'^2 - m^2)^{1/2}$$

Where $p_T$ and $p_z$ are respectively the transverse and longitudinal components of the momentum of the particle in the overall rest frame of the hadronic fireball. Similarly $\beta_T$ and $\beta_z$ are respectively the transverse and longitudinal components of the expansion velocity ($\beta$) of a hadronic fluid element. We assume that the expanding hadronic fluid element does not have any amount of whirl velocity component (i.e. azimuthal component) hence we have $\beta_\phi = 0$. The energy and the longitudinal momentum of the particle in terms of rapidity and transverse mass are given as

$$E = m_T \cosh y \qquad p_z = m_T \sinh y.$$

The transverse velocity component of the hadronic fireball, i.e. $\beta_T$, is assumed to vary with the transverse coordinate $r$ as [3,19]

$$\beta_T(r) = \beta_T^s \left(\frac{r}{R}\right)^n \tag{6}$$

where *n* is an index which fixes the profile of $\beta_T(r)$ and $\beta_T^s$ is the hadronic fluid surface transverse expansion velocity (i.e. when r = R) and is fixed in the model by using the following parameterization [4]

$$\beta_T^S = \beta_T^0 \sqrt{1 - \beta_z^2} \tag{7}$$

The above relation (or restriction) is also required to ensure that the velocity $\beta$ of any fluid element must satisfy

$$\beta = \sqrt{\beta_T^2 + \beta_z^2} < 1 \tag{8}$$

We also need to parameterize *R* [4,18] as,

$R = r_0 \; exp(-\frac{z^2}{\sigma^2})$, where $r_0$ is a model parameter which fixes the transverse size of the hadronic matter while $\sigma$ fixes the width of the matter distribution in the transverse direction[4]. The evidence of this requirement comes from the rapidity data which cannot be reproduced by using the same transverse size (or weight factor) for all fluid elements distributed along rapidity axis.

In the earlier analysis [18], $y_0$ was treated as a free running variable having no known dependence on the spatial z-coordinate within the fireball. In the present analysis, we choose a simple dependence of linear type where $y_0 \propto z$. This also ensures that under the transformation $z \to -z$, we will have $y_0 \to -y_0$, thereby preserving the symmetry of the hadronic fluid flow about z = 0 (or $y_0$ = 0) along the rapidity (or the z) axis in the centre of mass frame of the colliding nuclei. This is also consistent with the picture presented above leading to the above equations. Thus writing $y_0 = cz$, we obtain the following expression for the longitudinal velocity component of the hadronic fluid element:

$$\beta_z(z) = 1 - \frac{2}{exp(2cz)+1} \tag{9}$$

In the above equation when $z \to 0$ we obtain $\beta_z(z) \to 0$ and $\beta_T^s \to \beta_T^0$, while for $z \to \infty$ we obtain $\beta_z(z) \to 1$ and $\beta_T^s \to 0$. In our calculation we have used $c = 1$ as it provides a good fit to the overall data.

The contributions of various heavier hadronic resonances which decay, after the freeze-out has occurred, are also taken into account. The invariant spectrum of a *given* decay product of a *given* parent hadron in the local rest frame of a hadronic fluid element can be written as [4, 5, 20]:

$$E' \frac{d^3 n^{decay}}{d^3 p'} = \frac{1}{2p'} \left\{\frac{m_h}{p^*}\right\} \int_{E_-}^{E_+} dE_h \, E_h \left\{\frac{d^3 n_h}{d^3 p_h}\right\} \tag{10}$$

Where the subscript *h* stand for the decaying (parent) hadron. The two body decay kinematics gives the **product** hadron's momentum and energy in the "rest frame of the decaying hadron" as :

$$p^* = (E^{*2} - m^2)^{1/2} \tag{11}$$

$$E^* = \frac{m_h^2 - m_j^2 + m^2}{2m_h} \tag{12}$$

where $m_j$ indicates the mass of the *other* decay hadron produced along with the first one (under consideration having mass $m$). Thus the limits of integration in the Eq. (10) are:

$$E_\pm = \left\{\frac{m_h}{m^2}\right\} \{E'E^* \pm p'p^*\} \tag{13}$$

The $E'$ ($E_h$) and $p'$ ($p_h$) are, respectively, the product and decaying (parent) hadron's energy and momentum in the local *rest frame of the hadronic fluid element*. Considering Boltzmann type distribution for the decaying (parent) hadron in the *local rest frame of the hadronic fluid element* we can obtain the following expression for the invariant cross section of the product hadron

$$E' \frac{d^3 n}{d^3 p'} =$$

$$\frac{1}{2p'} \left\{\frac{m_h}{p^*}\right\} \lambda_h\, g_h\, e^{-\alpha\theta E'E^*} \left\{\frac{\alpha}{\theta} [E'E^* \sinh(\alpha\theta p'p^*) - p'p^* \cosh(\alpha\theta p'p^*)] + T^2 \sinh(\alpha\theta p'p^*)\right\} \tag{14}$$

where $\alpha$ and $\theta$ are given by $m_h/m^2$ and $1/T$, respectively.

In our analysis we have also applied the criteria of exact strangeness conservation. It is done in such a way that the net strangeness is conserved not only on the overall basis but also in every region of the fireball separately as they maintain different baryon chemical potentials along the z-axis. This is essential because as the rapidity of these regions of the fireball increases along the rapidity axis the baryon chemical potential ($\mu_B$) also increases, in accordance with Eq. (4). Hence the required value of the strange chemical potential ($\mu_S$) will accordingly vary with $\mu_B$ for a given value of temperature T. Consequently the value of the strange chemical potential ($\mu_S$) will vary with $y_0$.

## 3. Results and Discussion

### 3.1 Transverse momentum distributions

In an earlier analysis [18] the width of the distribution, σ, was determined from pion rapidity distribution as this was more sensitive to the value of sigma and less to the variation of baryon chemical potential. In our study the width of the distribution determined by the parameter $\sigma$ is chosen through the best fit of the given hadron's rapidity spectrum. The rapidity distribution is found to be sensitive to the values of the baryon chemical potential along the rapidity (collision) axis and the temperature. The model parameters *a, b, σ* and *T* therefore have a significant effect on the rapidity spectra. The flow velocity profile (or the given index *n*) has almost negligible effect on the rapidity spectra and a somewhat weak effect on the transverse momentum spectra of the hadrons in the range *n* = 1 – 2. The value of index *n* in our analysis is fixed to be unity. The parameter $\beta_T^0$ also has a weak effect on the rapidity spectra. Contrary to this the transverse momentum spectra of hadrons are very sensitive to $\beta_T^0$ as well as *T*. In our analysis we have therefore fixed the values of the parameters T and $\beta_T^0$ at freeze-out by obtaining a best fit to a given hadron's transverse momentum spectra, while the remaining parameters i.e. *a, b* and $\sigma$ are fixed by obtaining a best fit to its rapidity distribution (dN/dy) owing to its sensitivity to these parameters.

In figure 1, we have shown the transverse momentum distributions for protons and antiprotons. Using our model a best fit to the experimental data (shown by the solid circles) for Au+Au collision at $\sqrt{s_{NN}}$ = 200 GeV is obtained. The experimental data points are from the PHENIX collaboration taken for various hadrons over a wide range of $p_T$ [21]. The theoretical curves obtained by fitting the experimental data are shown by the solid curve. The transverse momentum distribution of the protons are best fitted with the thermal freeze-out conditions having $\beta_T^0$ = 0.66, T = 162 MeV while the remaining parameters have values σ = 4.3, a = 22.4 MeV and b = 9.1 MeV. For the antiprotons these values turn out to be $\beta_T^0$ = 0.67, T = 163 MeV, σ = 4.20, a = 22.4 MeV and b = 9.1 MeV. The values of the freeze-out parameters for protons and antiprotons are very similar thereby indicating a near simultaneous freeze-out for the protons and antiprotons in the dense hadronic medium. The minimum $x^2/DoF$ for the above two cases are 1.45 and 3.93, respectively.

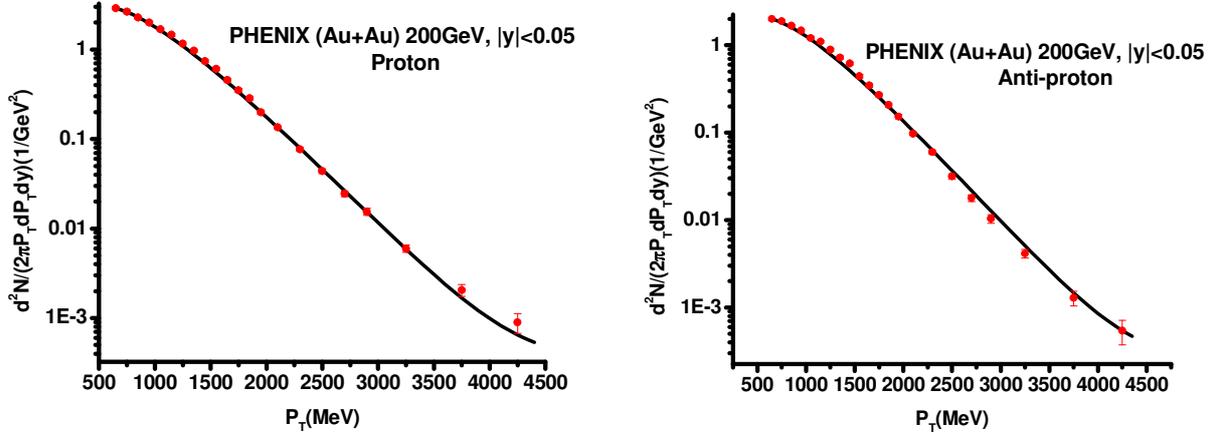

**Fig.1: Transverse momentum distribution of protons (left-panel) and anti-protons (right-panel)**

Next we analyze the transverse momentum ($p_T$) freeze-out spectra of the strange particles (mesons as well as singly and doubly strange hyperons) produced in the collisions of Au + Au at $\sqrt{s_{NN}} = 200$ GeV. The PHENIX data on the transverse momentum distributions of strange mesons (Kaons and anti-Kaons) are fitted using the same model. In figure 2, we have shown the transverse momentum distributions of Kaons ($K^+$) and anti-kaons ($K^-$). The freeze-out parameter values for Kaons are $\beta_T^0 = 0.56$, T = 161 MeV, $\sigma$ = 4.10, a = 22.4 MeV and b = 9.1 MeV, while for the anti-kaons these values turn out to be $\beta_T^0 = 0.58$, T = 161 MeV, $\sigma$ = 4.36, a = 22.4 MeV and b = 9.1 MeV. The minimum $x^2/DoF$ for the two cases are 5.91 and 1.62, respectively. The transverse flow parameter ($\beta_T^0$) values of the lighter particles (i.e. $K^+$ and $K^-$) are found to be smaller than those for the heavier particles (i.e. p, $\overline{p}$, $\Lambda$, $\overline{\Lambda}$ etc;).

This suggests that lighter particle's transverse momentum distributions are less influenced by the expansion of the hadronic fluid than the other heavier species present in it.

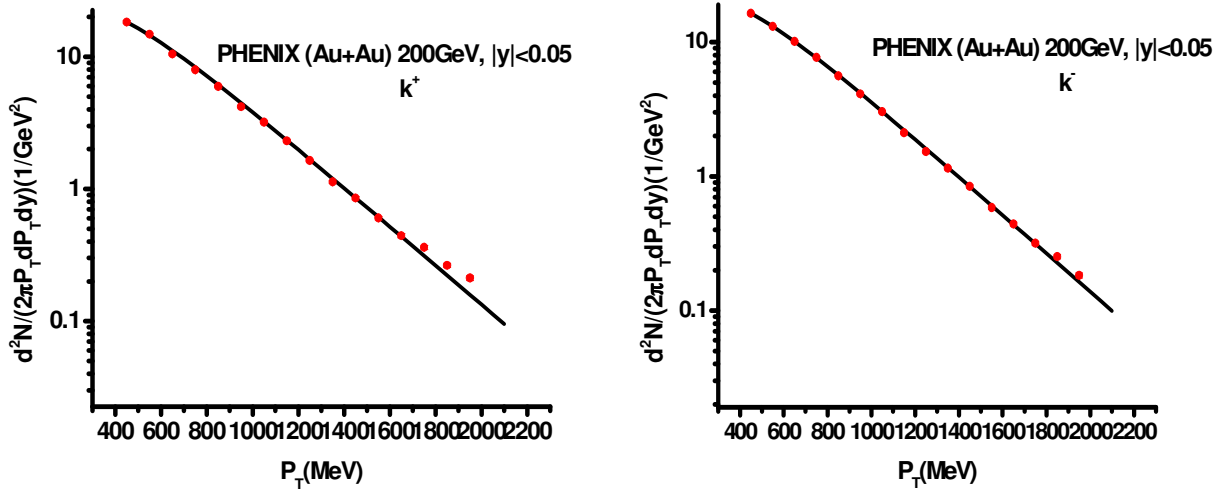

**Fig.2: Transverse momentum distribution of Kaons (left-panel) and anti-Kaons (right-panel)**

The transverse momentum spectra of the singly, doubly and triply strange hyperons are also fitted using the same model. These spectra show a higher freeze-out temperature than other particles (protons and Kaons). The increase in temperature for strange particles is an indication of their somewhat early freeze-out and hence their spectra exhibit larger thermal temperatures. The transverse momentum distribution for $\Lambda$ and $\overline{\Lambda}$ are shown in the figure 3. Both $\Lambda$ and $\overline{\Lambda}$ are fitted with the freeze-out conditions of collective flow $\beta_T^0 = 0.60$ and kinetic/thermal temperature of T=167 MeV, while the values of the other parameters are $\sigma = 4.25$, a = 22.0 MeV and b = 9.1 MeV. The minimum $x^2/DoF$ for the two cases are 2.05 and 2.46, respectively.

The transverse momentum distributions for $\Xi^-$ is also well described with the freeze-out condition of $\beta_T^0 = 0.60$, T=186 MeV, $\sigma = 4.25$, a = 22.4 MeV and b = 9.1 MeV. The minimum $x^2/DoF$ for this case is 0.40. The $\overline{\Xi^-}$ transverse momentum distribution is fitted by using $\beta_T^0 = 0.59$, T=188 MeV, $\sigma = 4.25$, a = 22.4 MeV and b = 9.1 MeV. The minimum $x^2/DoF$ for this case is 0.72. In Figure 4, we have shown the transverse momentum distributions of $\Xi^-$ and $\overline{\Xi^-}$. The freeze-out conditions extracted from the transverse momentum distribution of $(\Omega^- + \Omega^+)$ are found to be nearly same as those for the $\Xi^-$ and $\overline{\Xi^-}$. The minimum $x^2/DoF$ for this case is 2.10.

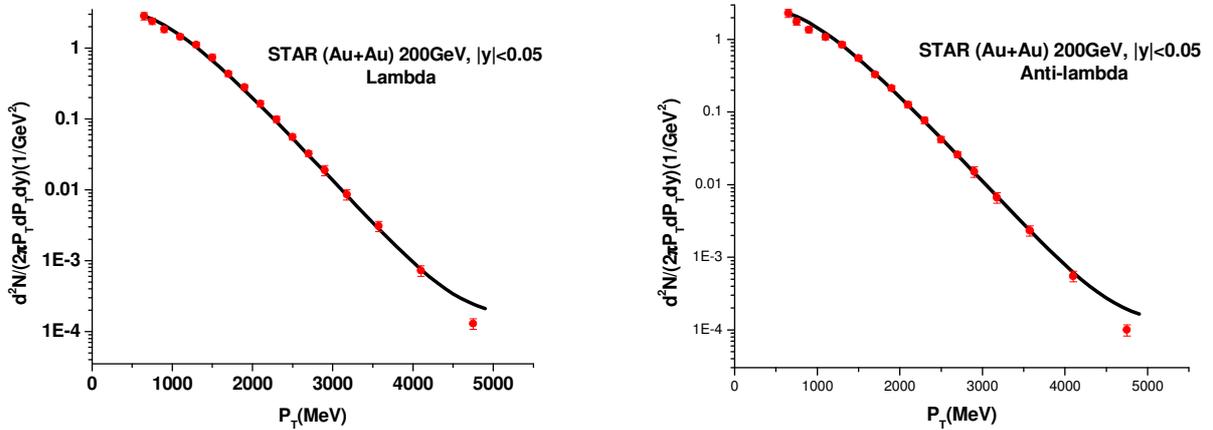

**Fig.3: Transverse momentum distribution of lambda (left-panel) and anti-lambda (right-panel)**

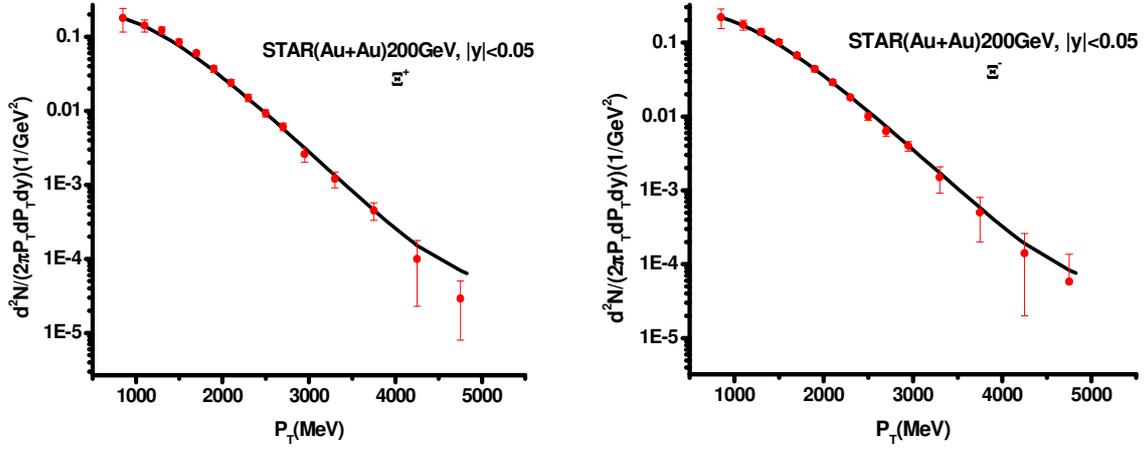

**Fig.4: Transverse momentum distribution of cascade (left-panel) and anti-cascade (right-panel)**

In Figure 5, we have shown the transverse momentum distributions of $(\Omega^- + \Omega^+)$. The experimental data points are shown by solid circles in all these graphs while the solid curves represent out theoretical results .

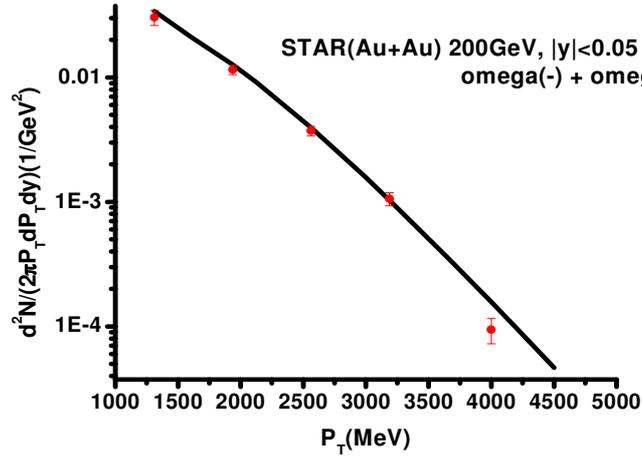

**Fig.5: Transverse momentum distribution of $(\Omega^- + \Omega^+)$**

## 3.2 Rapidity Distributions

The available rapidity distribution data of p, $\bar{p}$, $\bar{p}/p$, $K^+$, $K^-$, $\bar{\Lambda}/\Lambda$ and $\bar{\Xi}/\Xi$ ratios are shown in the figures 6 to 9. The available data are from BRAHMS collaboration at $\sqrt{s_{NN}} = 200$ GeV for (0-5)% most central Au+Au collisions. As discussed above the experimental rapidity distributions data are best fitted with the same freeze-out conditions as their corresponding transverse momentum spectra. The minimum $x^2/DoF$ are obtained by using the same temperature and the transverse expansion flow parameter $\beta_T^0$ values. The $x^2/DoF$ is minimized with respect to the variables a, b and σ which provide satisfactory values of $x^2/DoF$. These values for the cases of p, $\bar{p}$, $\bar{p}/p$, $K^+$ and $K^-$ are 2.30, 2.89, 0.73, 3.03 and 1.58, respectively. For the case of $\bar{\Lambda}/\Lambda$ and $\bar{\Xi}/\Xi$ we have only one experimental data point available at the mid-rapidity.

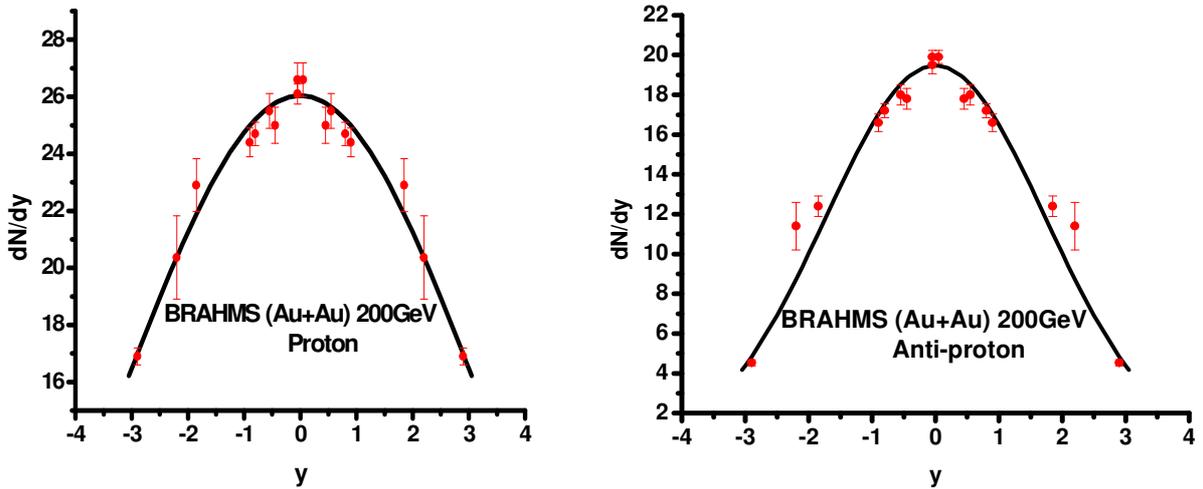

**Fig. 6:** Rapidity distribution of protons (left-panel) and anti-protons (right-panel). The experimental data are for the 0 – 5% most central collisions

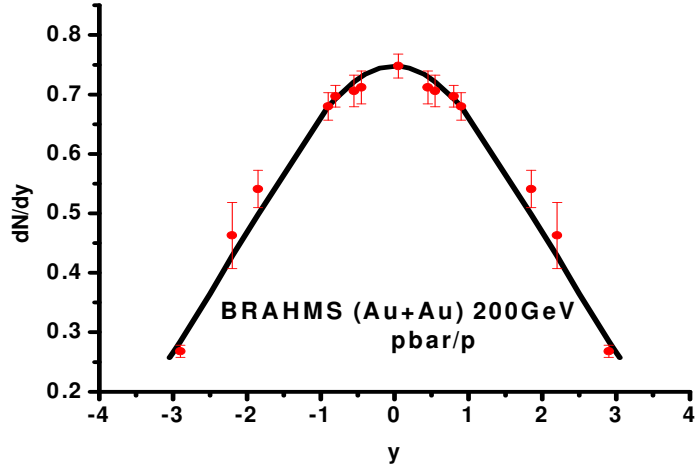

**Fig.7: Rapidity distribution of $\bar{p}/p$ ratio.**

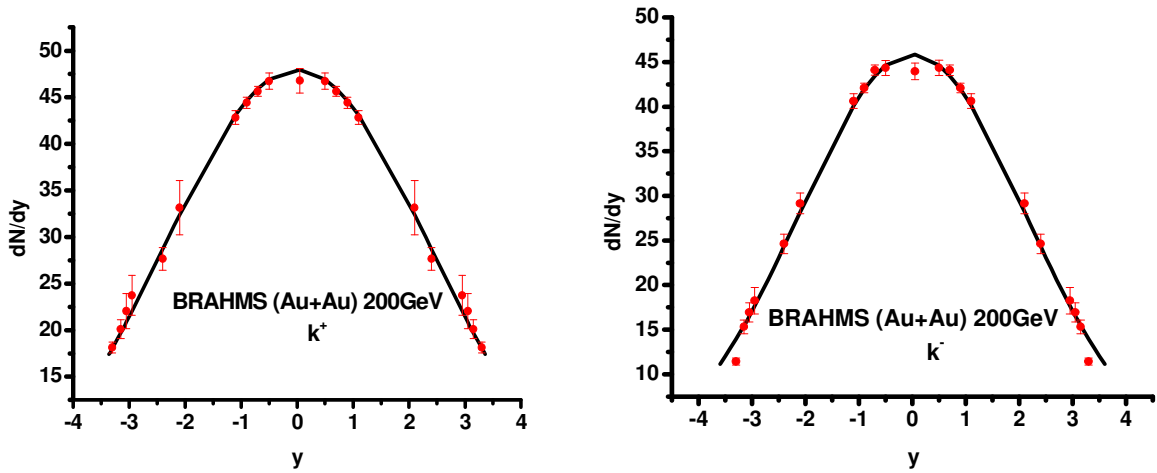

**Fig. 8: Rapidity distribution of Kaons (left-panel) and anti-Kaons (right-panel). The experimental data are for the 0 – 5% most central collisions**

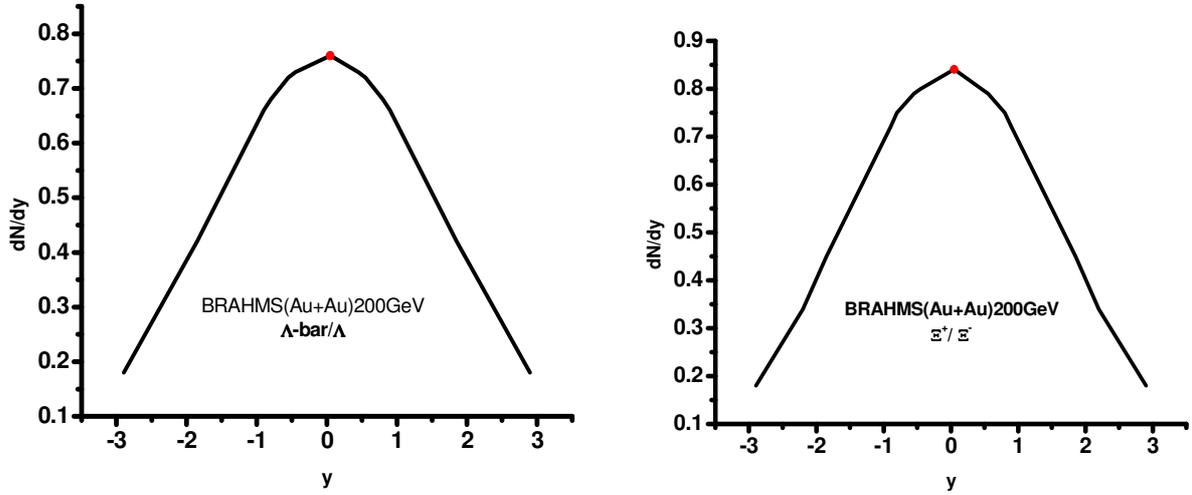

**Fig.9:** Rapidity distributions of $\overline{\Lambda}/\Lambda$ ratio (left-panel) and $\overline{\Xi}/\Xi$ (right-panel). The experimental data are for the 0 – 5% most central collisions

As a comparison in figures 10 and 11 we also show the UrQMD calculation results [22] which attempt to describe the rapidity distributions of protons (& anti-protons) and Kaons (and ant-Kaons), respectively, at various energies including the highest RHIC energy i.e, $\sqrt{s_{NN}} = 200$ GeV. We see that in the case of proton's rapidity distribution the calculated UrQMD result exhibits a trend which is actually opposite to the experimentally obtained behavior. In case of antiproton, Kaon and anti-Kaon rapidity distributions, the calculated results highly under predict the experimental data at and near mid-rapidity regions, while for larger rapidities (>2) it over predicts the experimental data. In contrast to it we have successfully described not only the rapidity but also the transverse momentum distributions, simultaneously, of the various hadronic species in the framework of single statistical thermal freeze-out model in the entire rapidity and $p_T$ range.

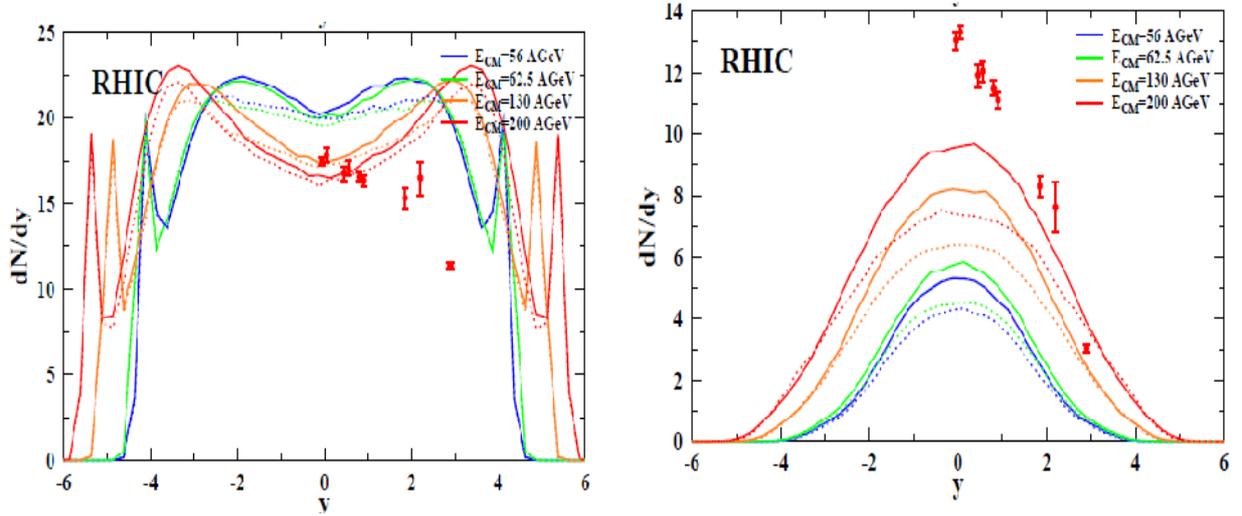

Fig.10: Rapidity spectra of protons (left panel) and anti-protons (right panel) for central Au+Au/Pb+Pb collisions. UrQMD-2.3 calculations are depicted with full lines, while UrQMD-1.3 are depicted with dotted lines. The corresponding data from different experiments are depicted with symbols [22].

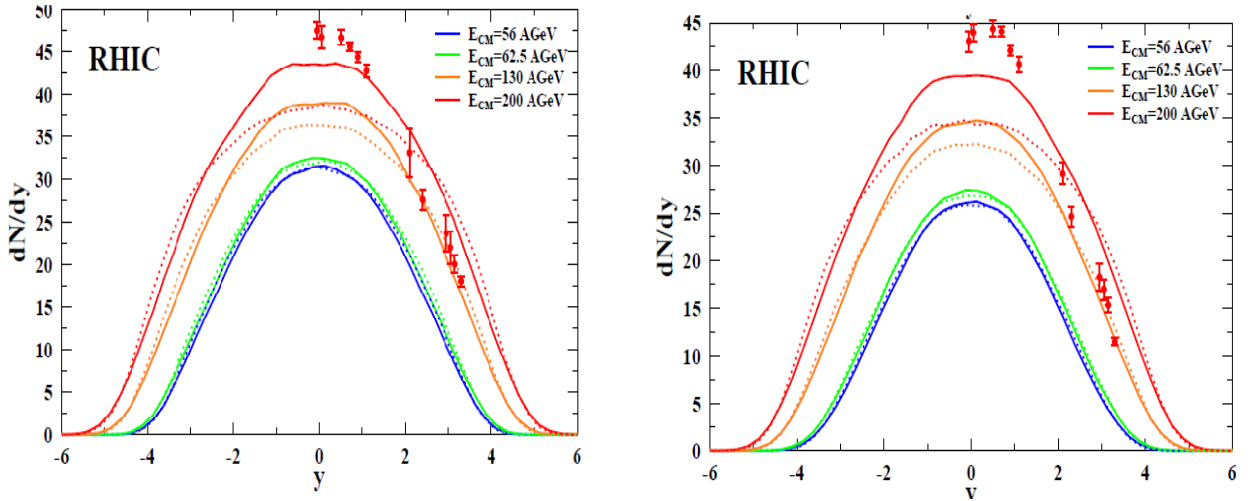

Fig.11: Rapidity spectra of Kaons (left panel) and anti-Kaons (right panel) for central Au+Au/Pb+Pb collisions. UrQMD-2.3 calculations are depicted with full lines, while UrQMD-1.3 are depicted with dotted lines. The corresponding data from different experiments are depicted with symbols [22].

## 4 Summary and Conclusion

We have utilized our thermal model to successfully describe the rapidity *as well as* transverse momentum spectra of various hadrons produced in the 0 – 5% most central Au + Au collisions at $\sqrt{s_{NN}} = 200$ GeV at RHIC. This is achieved in our model by simultaneously incorporating a longitudinal as well as a transverse flow effect of the expanding dense hadronic matter. The system's transverse size is assumed to decrease following a Gaussian profile in the z-coordinate. The model provides a very good description of the transverse momentum spectra of $\Xi, \bar{\Xi}, \Lambda, \bar{\Lambda}, p, \bar{p}, (\Omega^- + \Omega^+), K^+, K^-$ and the rapidity spectra of $p, \bar{p}, \bar{p}/p, K^+, K^-, \bar{\Lambda}/\Lambda$ and $\bar{\Xi}/\Xi$. The freeze-out conditions of various hadronic species are determined. It is found that the singly, doubly and triply strange hyperons freeze-out conditions corresponds to higher thermal freeze-out temperature as compared to the non-strange baryons and lighter strange mesons. Our results provide a much better description of these data than those obtained from the UrQMD.

## 5. Acknowledgments

Riyaz Ahmed is grateful to Council of Scientifc and Industrial Research, New Delhi for awarding Senior Research Fellowship. Inam-ul Bashir is thankful to University Grants Commission for providing Basic Scientific Research (BSR) Fellowship.

# 6. References

[1] L. P. Csernai, Y.Cheng, V. K. Magas, I. N. Mishustin, D. Strottman, *Nucl. Physics* **A 834**, 261 (2010).

[2] V.K. Magas, L.P. Csernai, D.D. Strottman, Phys. Rev. **C 64** (2001) 014901; *Nucl. Phys*. **A712** (2002) 167; E. Kornas *et al.* NA 49 Collaboration, *Eur. Phys. J.* **C49** (2007) 293

[3] Saeed Uddin, N. Akhtar and M. Ali, *Int. J. Mod. Phys*. **21**, 1472 (2006)

[4] Saeed Uddin, Jan Shabir Ahmad, Waseem Bashir and Riyaz Ahmad Bhat. *J. Phys.* **G 39**, 015012 (2012)

[5] Saeed Uddin *et al* 2010 *Acta Physica Polonica* **B Vol. 41 (2010).**

[6] V.P. Kondrat'ev, G. A. Feofilov, 2011, published in Fizika Elementarnykh Chastits i atomnogo yadra, 2011, vol.42, No. 6.

[7] Adler C et al STAR Collaboration Phys. Rev. Lett. 87, 262302, (2001).

[8] Becattini F et al Phys. Rev. C 64, 024901, (2001).

[9] Braun-Munzinger P, Magestro D, Redlich K and Stachel J 2001 *Phys.Lett. B* 518 41

[10] Florkowski W, Broniowski W and Michalec M 2002 *Acta Phys.Pol.B* 33 761

[11] L.A. Stiles and M. Murray,*nucl-ex*/0601039

[12] J.Cleymans et al.*arxiv*:0712.2463v4 [hep-ph] 2008

[13] J. Cleymans, S. Kabanab, I. Krausc, H. Oeschlerd, K. Redlichf, N. Sharma *arxiv*:1107.0450v1 [hep-ph] 3 Jul 2011

[14] F.Becattini and J. Cleymans, J. Phys. **G34**(2007) S959

[15] F.Becattini et al., Proceedings of Science, CPOD07(2007) 012

[16] W. Broniowski and B. Biedron J.Phys .**G35**(2008) 044018

[17] Sarkar S, Satz H Sinha B,The Physics of the Quark Gluon Plasma, Introductory Lectures, Lect. Notes Physics 785 (Spinger, Berlin, Heidelberg 2010).

[18] F.Becattini, J. Cleymans and J. Strmpfer, *arXiv*; 0709.2599 v1[hep-ph]

[19] Koch P, Muller B and Rafelski J 1986 *Phys. Rep.* **142** 187.

[20] Hagedorn R 1964 *Relativistic Kinematics* (New York: Benjamin)

[21] Adler S S et al. (PHENIX Collaboration), Preprint: nucl-ex/0307022.

[22] Hannah Peterson, Marcus Bleicher, Steffen A. Bass and Horst Stocker, Preprint: arXiv:0805.0567v1 [hep-ph], 2008.